\documentclass[3p,twocolumn]{elsarticle}

\usepackage{lineno,hyperref,amsmath}
\modulolinenumbers[5]

\newcommand{\singlePlotSize}{0.95\columnwidth}

\journal{Nuclear Instruments and Methods in Physics Research Section A}

\begin{document}

\begin{frontmatter}

\title{Novel proposal for a low emittance muon beam using positron beam on target}


\author[frascati]{M. Antonelli \corref{mycorrespondingauthor}}
\cortext[mycorrespondingauthor]{Corresponding author}

\author[frascati]{M. Boscolo}
\author[frascati]{R. Di Nardo}
\author[esrf]{P. Raimondi}

\address[frascati]{INFN/LNF, Via Enrico Fermi 40, 00044 Frascati (Roma), Italy}
\address[esrf]{ESRF, Grenoble, France}

\begin{abstract}
  Muon beams are customarily obtained via $K/\pi$ decays produced in proton interaction on target.
  In this paper we investigate the possibility to produce low emittance muon beams
  from electron-positron collisions at centre-of-mass energy just above the $\mu^{+}\mu^{-}$ production threshold
  with maximal beam energy asymmetry, corresponding to a positron beam of about
  45 GeV interacting on electrons on target. We present the main features of this scheme
  with an outline of the possible applications.
\end{abstract}


\end{frontmatter}


\section{Introduction}

Muon beams are customarily obtained via $K/\pi$ decays produced in proton interaction on target.
Their use in high energy physics experiments has a continuous increasing interest
for rare decays searches, precision measurement experiments, neutrino physics and for
muon colliders feasibility studies. Several dedicated experiments are ongoing
to produce high intensity muon beams with low emittance. 
In this paper we will investigate the possibility to produce
low emittance muon beams from a novel approach, using the
electron-positron collisions at centre-of-mass energy just above the $\mu^{+}\mu^{-}$  production
threshold with maximal beam energy asymmetry, that corresponds to about 45 GeV
positron beam interacting on an electron target. Previous studies on this subject are reported in ref.~\cite{Snowmass,Snowmass1}.
Our proposal is simpler with respect to present conventional projects where muons are produced by a proton source.
One important aspect is that in our proposal muon cooling would not be necessary. 
The most important key properties of the muons produced by the positrons on target are:
\begin{itemize}
  \item the low and tunable muon momentum in the centre of mass frame
  \item large boost, being about $\gamma\sim$200.
\end{itemize}
These characteristic results in the following advantages:
  \begin{itemize}
  \item the final state muons are highly collimated and have small emittance;
  \item the muons have an average laboratory lifetime of about 500 $\mu$s.
\end{itemize}
In the  section \ref{sec:2} we describe the main processes at
the energy of interest. In section two we describe the key issue
of the options on the target. The value of the $e^{+}e^{-}\rightarrow\mu^{+}\mu^{-}$ cross
section is of about 1 $\mu$b just above threshold, requiring a target with very high electron
density to obtain a reasonable muon production efficiency.
We estimate the requirement on the electron density on the target
above $10^{20}$ electrons/cm$^{-3}$. Such high-density values can be obtained
either in a liquid or solid target or, possibly, in a more exotic
solution like in crystals. We discuss the solid target solution
and the crystals. Also a plasma excited via a syncronized
electron beam could be a solution. 
In section \ref{sec:3} we discuss our first estimates that show how this option does not seem practicable.
 However, we think that that further studies are worth to be performed on this option.
Studies of the positron source will be reviewed in section \ref{sec:4},
followed, in section \ref{sec:5}, by the schemes of the muon production.
Finally, in section \ref{sec:6}, rate and beam properties estimates for muon
collider that can be obtained with our proposal are given.
We conclude with a proposal for a design study, especially
for the key innovative aspects  that need to be investigated.

\section{Processes at $\sqrt{s}$ around 0.212 GeV}
\label{sec:2}
The dominant processes at $\sqrt{s}$ around 0.212 GeV are mainly three:
\begin{enumerate}
  \item $e^{+}e^{-}\rightarrow\mu^{+}\mu^{-}$ : $\mu^{+}\mu^{-}$ production
  \item $e^{+}e^{-}\rightarrow e^{+}e^{-}\gamma$ : Bhabha scattering
  \item $\gamma\gamma$ scattering.
\end{enumerate}
Our goal is to maximize the muons production and muon beam parameters.
 In the following sub-section \ref{subsec:2.1} we analyse their 
dependencies and features. The second and third process are instead
side effect that reduce the efficiency of the first one.
We will discuss in the  sub-section  \ref{subsec:2.2} the second process.
They have been simulated with the {\texttt{BabaYaga}} event generator \cite{CarloniCalame:2003yt} with the
exception of the collinear radiative Bhabha scattering, for which we used
{\texttt{BBBrem}} \cite{Kleiss:1994wq}. The last process, the  $\gamma\gamma$ scattering, will not be discussed
in detail, having a cross section that is smaller than that for the $\mu^{+}\mu^{-}$  production.
We note that, if needed, the method proposed here for the muons production could
also allow the production of high energy collimated photons.

\subsection{The Process $e^{+}e^{-}\rightarrow\mu^{+}\mu^{-}$ }
\label{subsec:2.1}
\begin{figure}[htbp!]
  \centering
  \includegraphics[width=\singlePlotSize]{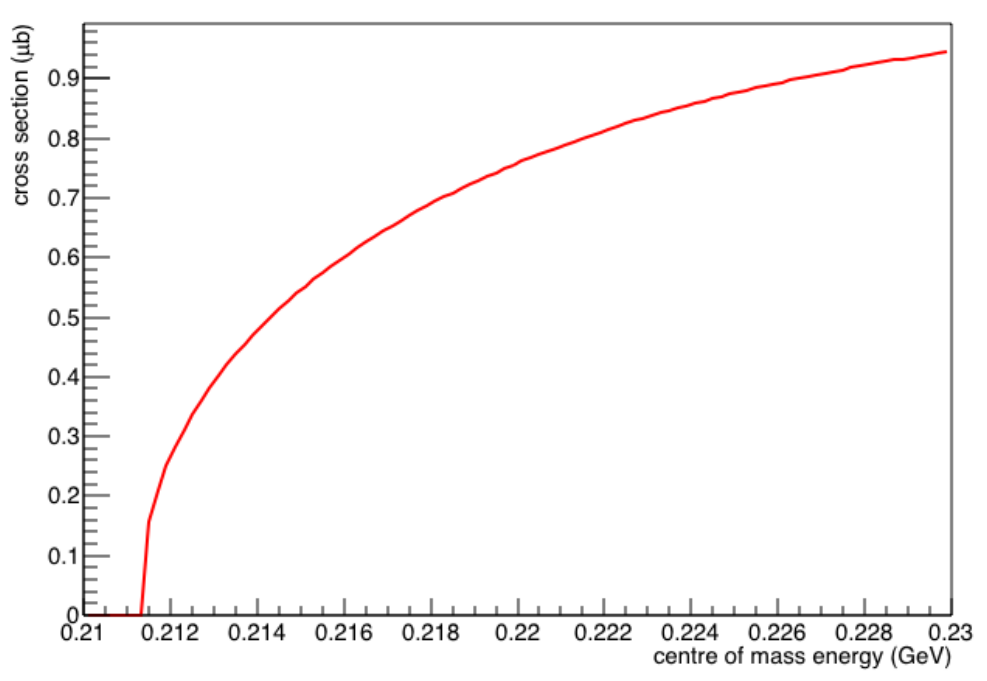}
  \caption{Cross section as a function of  $\sqrt{s}$ of the $e^{+}e^{-}$ collisions.
    \label{fig:xsec}}
\end{figure}

We discuss here the dependencies that can maximize the muons production
and at the same time also minimize the muon bunch emittance and energy spread,
when required. The main parameters that play a role are: the dependence
of the scattering angle distribution of the outgoing muons, the
muons energy distribution and the cross section on the centre-of-mass energy.
The cross section for continuum muon pair production $e^{+}e^{-}\rightarrow\mu^{+}\mu^{-}$ just above threshold
is obtained using the Born cross section, enhanced by the Sommerfeld-Schwinger-Sakharov (SSS)
threshold Coulomb resummation factor \cite{Brodsky:2009gx}. The value of this cross section
is shown in figure~\ref{fig:xsec} as a function of the centre-of-mass energy. In this
figure we see that the cross section approaches its maximum value of about 1$\mu$b at $\sqrt{s} \sim$0.230 GeV.

\begin{figure}[htbp!]
  \centering
  \includegraphics[width=\singlePlotSize]{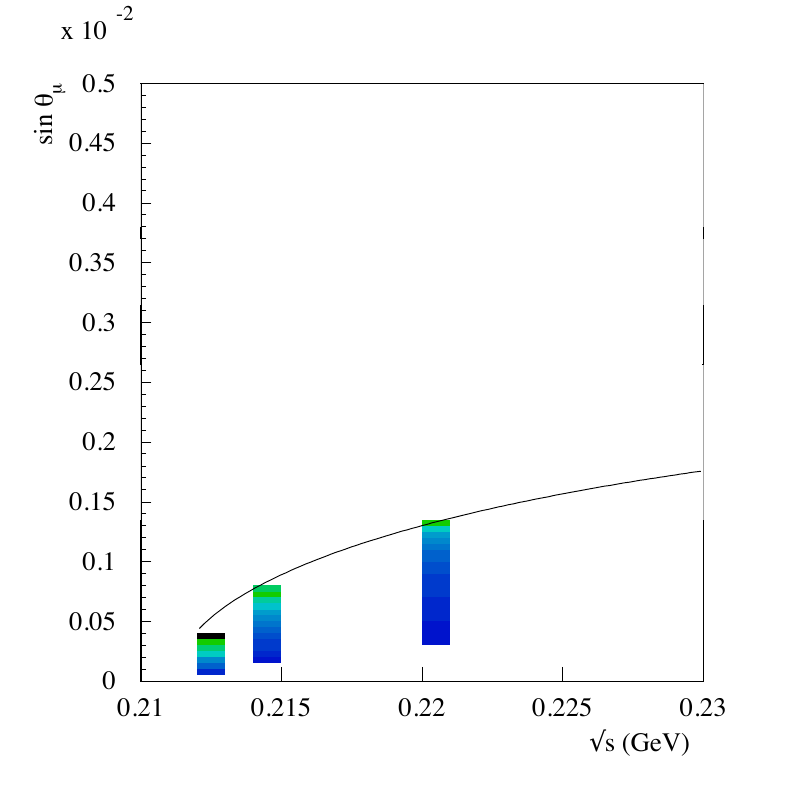}
  \caption{Muon scattering angle distribution as a function of  $\sqrt{s}$  of the $e^{+}e^{-}$  collisions.
    \label{fig:muang}}
\end{figure}

In our proposal these values of  $\sqrt{s}$  can be obtained from fixed target interactions with a positrons beam energy of 
    $$
    E_{+}\approx  s/(2 m_{e}) \approx 45 ~\hbox{GeV} 
    $$
where $m_{e}$  is the electron mass, with a boost of $\gamma\approx  E_{+}/\sqrt{s} \approx \sqrt{s}/(2 m_{e}) \approx 220$.
The scattering angle of the outcoming muons $\theta_{\mu}$  is maximum for the muons emitted
orthogonally to positron beam (in the rest frame) and its value depends on $\sqrt{s}$ (see figure~\ref{fig:muang}).

In the approximation of $\beta_{\mu}=1$, where $\beta_{\mu}$  is the muons velocity, one can easily obtain for the maximum scattering angle:
\begin{equation}
  \theta_{\mu}^{max}= \frac{4m_{e}}{s} \sqrt{\frac{s}{4}-m_{\mu}^{2} }
    \end{equation}

The value of the scattering angle $\theta_{\mu}$   increases with the $\sqrt{s}$   with approximately
the same shape as the cross section of the $\mu^{+}\mu^{-}$ production.
The difference between the maximum and the minimum energy of the muons produced
at the positron target ($\Delta E_{\mu}$) also depends on  $\sqrt{s}$,  and with the  $\beta_{\mu}=1$ approximation we get:
\begin{equation}
\Delta E_{\mu} =\frac{\sqrt{s}}{2m_{e}} \sqrt{\frac{s}{4}-m_{\mu}^{2}}
\end{equation}

\begin{figure}[htbp!]
  \centering
  \includegraphics[width=\singlePlotSize]{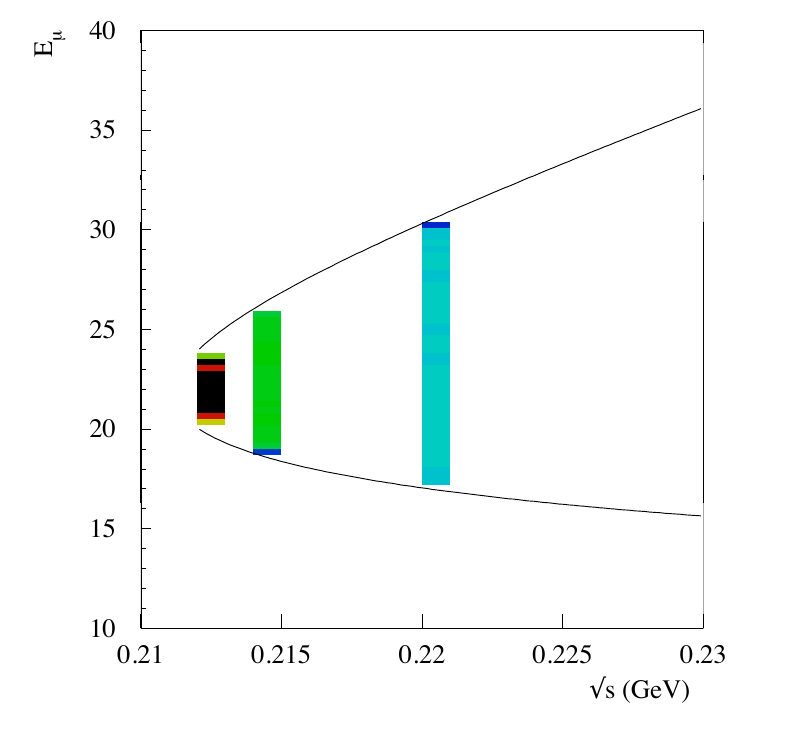}
  \caption{Muons energy distribution as a function of $\sqrt{s}$.
    \label{fig:muene}}
  \end{figure}

\begin{figure}[htbp!]
  \centering
  \includegraphics[width=\singlePlotSize]{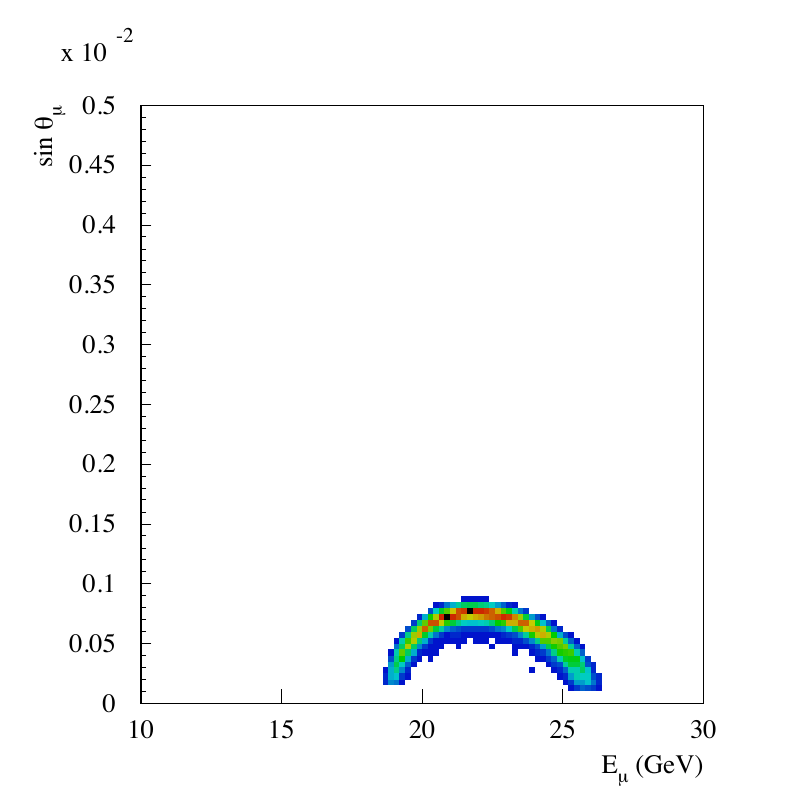}
  \caption{Scattering angle vs muons energy distribution for the $\sqrt{s}$=0.214 GeV  case
    \label{fig:mu_scatvale}}
\end{figure}

These values have to be folded with the muons angular distribution in the rest
frame, that is: $(1+\cos^{2}\theta_{\mu}^{*})$, where $\theta_{\mu}^{*}$   is the muon scattering angle in the rest frame.
The value of $\sqrt{s}$   has to be optimized to maximize the $\mu^{+}\mu^{-}$ production
and to minimize the beam angular divergence and eventually the energy spread.
The $\theta_{\mu}$  distribution obtained with the {\texttt{BabaYaga}} generator is shown in
figure~\ref{fig:muang} for different $\sqrt{s}$   values. Muons produced with very small momentum
in the rest frame are well contained in a cone of about  $5\cdot 10^{-4}$ rad for   $\sqrt{s}$=0.212 GeV,  the cone
size increases to $\sim 1.2\cdot 10^{-4} $  rad at  $\sqrt{s}$=0.220 GeV. Similarly, the energy
distribution of the muons, as shown in figure ~\ref{fig:muene}, has an $RMS$ that increases with $\sqrt{s}$, from about
1 GeV at $\sqrt{s}$=0.212 GeV to ~3 GeV at  $\sqrt{s}$=0.220 GeV. The muons beam energy has
the typical correlation with the muons emission angle as shown in figure ~\ref{fig:mu_scatvale} for  $\sqrt{s}$=0.214 GeV.

Another solution in principle could be to produce the muonium below the
$\mu^{+}\mu^{-}$ threshold that can be eventually dissociated in the interaction with
the medium. It has been studied in Ref.~\cite{Brodsky:2009gx}, where it is shown that the $e^{+}e^{-}$  width is proportional to 1/n where n 
indicates the muonium energy level. The cross section for the $S^1$ state in the narrow width approximation is about:
$$
  10^{-9}  \rm{mb}  ~ ~E_{+}/ \sigma_{E_{+}} ,
$$

where $\sigma_{E_{+}}$ is the positrons beam energy spread. This value implies
that the use of this process for copious muons production is not realistic.

\subsection{The process $e^{+}e^{-}\rightarrow e^{+}e^{-}\gamma$}
\label{subsec:2.2}

\begin{figure}[htbp!]
  \centering
  \includegraphics[width=\singlePlotSize]{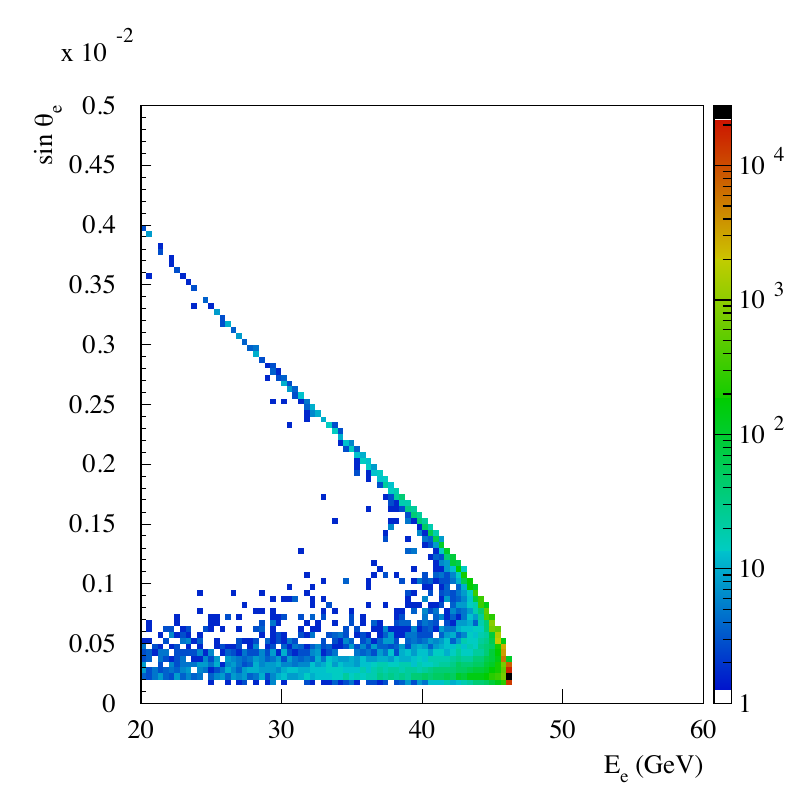}
  \caption{Distribution of the outgoing positrons scattering angle as a
    function of the energy of the positrons beam after
    its scattering on the target, for a positron beam energy before the scattering of $E_{+}=46$ GeV.
    \label{fig:angVsScatEnePos}}
\end{figure}

The Bhabha scattering represents the largest source of beam loss
in this study, setting an upper limit on the muons
production from positrons on target.
The large angle case has been simulated in the rest frame using BabaYaga
with radiative photons energy, $E_{\gamma}^{*}<$10 MeV , and a scattering angle $\theta_{gamma}>10^\circ$.
The total cross section in the  region of $\sqrt{s}=$ 0.2 GeV is
$$
  \sigma_{Bhabha}\approx 0.6 ~\hbox{mb}.
$$
As expected, the process proceeds via t-channel
and most of the generated events are produced at a very small positrons
scattering angle $\theta_{+}$. Figure~\ref{fig:angVsScatEnePos} shows the distribution of the outgoing
positrons scattering angle as a function of the energy of the outgoing positrons, for a beam  positron
 energy of  $E_{+}=46$ GeV impinging the target.

The distribution indicates that the beam loss due to this
process can be substantially decreased with reasonable acceptances.
The largest part of the $e^{+}e^{-}$ cross section comes from the collinear
radiative Bhabha scattering. It has been simulated with \texttt{BBBrem}\cite{Kleiss:1994wq}.
The total cross section is about 150 mb for a  $E_{\gamma}>0.01E_{+}$
and it gets to about 60 mb for $E_{\gamma}>0.1E_{+}$. This process sets a limit to the $\mu$ pairs
production, as it sets limits to the beam lifetime in high  luminosity $e^{+}e^{-}$  colliders.

\section{Target options}
\label{sec:3}
The number of $\mu^{+}\mu^{-}$ pairs produced per positron bunch on target is:
\begin{equation}
\label{eq:nmu}
  n(\mu^{+} \mu^{-})=n^{+} \rho^{-} l \sigma(\mu^{+} \mu^{-})      
\end{equation}
where $n^{+}$ is the number of positrons in
the bunch, $\rho^-$ is the electron density in the medium, $l$ is
the thickness of the target, and $\sigma(\mu^{+} \mu^{-})$ is the muon pairs production cross section.
As described in the previous section, the dominant process at these energies
is the collinear radiative Bhabha scattering with a cross section of about of 150 mb
actually setting the value of the positron beam interaction length
for a given pure electron target density value. Using as reference value for the positron
beam degradation when its current is decreased by 1/$e$, i.e. one beam lifetime, one can
determine the maximum achievable value for the target density and length:
\begin{equation}
(\rho^- l)_{max} =  1/\sigma(rad.bhabha)\approx 10^{25}  \hbox{cm}^{-2}
\end{equation}
The ratio of the muon pair production cross section to the radiative
bhabha cross section determines the maximum value of the {\it muons conversion efficiency} $eff(\mu^+\mu^-)$,
that can be obtained with a pure electrons target.
In the following we will refer to $eff(\mu^+\mu^-)$ defined as the ratio of the number of produced $\mu^+\mu^-$ pair
to the number of the incoming positrons.
Easily one can see that  the upper limit of $eff(\mu^+\mu^-)$  is of the order of $10^{-5}$, so that:
\begin{equation}
 n(\mu^+ \mu^-)_{max}\approx n^+ 10^{-5}.
\end{equation}

\subsection{Plasma target option}
The option of the plasma target has been considered and studied in some details.
It is known that an enhanced electron density can be obtained at the border of the
blow-out region in excited plasma. This solution provides with good approximation
 an ideal electron target and will also benefit from a strong and continuous beam focalization thanks to the Pinch effect~\cite{Chen:1986zr}.
An enhancement of about $10^2$  in the number of electrons can be obtained
in a region of about 100 $\mu$m just before the blow-out, for a plasma with density of 
$n_p = 10^{16}  electrons/\hbox{cm}^3$ ~\cite{PriveteCom:priv}.
The size of the electrons high density region scales as 1/$n_p$ , such that at useful
positrons densities values in the order of  $n_p=O(10^{20}$) will be reached
in regions in the $\mu$m range, making the plasma option hardly practicable.

\subsection{Conventional targets option}
Electromagnetic interactions with nuclei are dominant in conventional targets.
In addition, no intrinsic focusing effects are expected in this case, thus setting limits
for the target thickness to not increase the muons beam emittance $\epsilon_{\mu}$.
Assuming a uniform distribution in the transverse $x-x'$ plane\footnote{actually the distribution has and exponential
  fall with  $e^{-z/\lambda}$ being  $z$  the longitudinal  target coordinate and $\lambda$
  the smaller interaction length among all involved processes.
  The approximation is valid for $\lambda<l$.}
the emittance contribution due to the target thickness is :
\begin{equation}
\epsilon_{\mu}=\frac{xx'^{max}}{12}=\frac{l(\theta_{\mu}^{max})^{2}}{12}
\end{equation}
The number of $\mu^+\mu^-$ pairs produced per crossing has the form given by the relation~\ref{eq:nmu},  with
\[
\rho=N_A/A \rho Z
\]
being $Z$ the atomic number, $A$ the mass number, $N_A$ the Avogadro constant and $\rho$ the material density.
In addition, the multiple scattering contributes to the emittance increase according to:
\[
x'_{RMS}\sim \frac{0.0136}{P(\hbox{GeV})} \sqrt{0.5 l}
\]
with $l$ expressed in radiations length unit and
\[
x_{RMS}\sim  x'_{RMS} 0.5 l \sqrt{3}
\]
The bremsstrahlung process governs the positrons beam degradation in this case and it scales with the radiation length.

On one side to minimize the emittance there is the need of a small length $l$,
on the other side compact materials have typically small radiation length causing
an increase of the emittance due multiple scattering and fast positron
beam degradation due to bremsstrahlung. The production efficiency is instead
proportional to the electrons density.
Positrons survival probability is also an issue to be considered not only for long targets
(as long as one radiation length: $l\sim X_0$) but also if the positron beam is
recirculated to increase the positron rate impinging the target.
Relevant properties of the materials considered in our study are given in Table~\ref{tab:tab1};
together with the atomic and mass numbers Z and A are reported also the radiation length $X_0$, the interaction
length $\lambda(\mu^+\mu^-)$ and the ratio $\lambda(\mu^+\mu^-)/X_{0}$, being inversely proportional
to the maximum value that can be obtained for  $eff(\mu^+\mu^-)$.

\begin{table*}[htbp!]
  \centering
  \caption{Relevant properties for some materials considered suitable for the
    target, $\lambda(\mu^{+}\mu^{-})$ has been calculated with a cross section of 1 mb. \label{tab:tab1}}
  \vspace{0.05cm}
  \begin{tabular}{*{5}{c}}
    \hline\hline
    \noalign{\vspace{0.05cm}}
    Z    &   A         &  $X_0$(cm) &    $\lambda(\mu^+\mu-)$(cm)   &      $\lambda(\mu^+\mu-)/X_{0}$ \\  
    \hline
    1    &   1.00794   &  900.6     &    2.4 $~10^{7} $                 & $    2.7 ~10^{4}  $                 \\
    2    &   4.0026    &  786       &    2.8 $~10^{7}   $               &   $  3.5 ~10^{4}   $                \\
    3    &   6.941     &  156.2     &    7.2 $~10^{6}    $              &    $ 4.6 ~10^{4}  $                 \\
    4    &   9.01218   &  35.2      &    2.0 $~10^{6}    $              &  $   5.7 ~10^{4}  $                 \\
    5    &   10.811    &  22.2      &    1.5 $~10^{6 }    $             &   $  6.8 ~10^{4}    $               \\
    6    &   12.0107   &  21.4      &    1.7 $~10^{6}   $               &  $   7.8 ~10^{4}   $                \\
    6    &   12.0107   &  19.3      &    1.5 $~10^{6}   $               &   $  7.8 ~10^{4}   $                \\
    6    &   12.0107   &  18.8      &    1.5 $~10^{6}   $               &   $  7.8 ~10^{4}   $                \\
    6    &   12.0107   &  12.1      &    9.4 $~10^{5}   $               &   $  7.8 ~10^{4}  $                 \\
    29   &   63.546    &  1.4       &    4.1 $~10^{5}   $               &   $  2.8 ~10^{5}   $                \\
    74   &   183.84    &  0.4       &    2.1 $~10^{5}  $                &   $  6.1 ~10^{5}  $                 \\
    82   &   207.2     &  0.6       &    3.7 $~10^{5}    $              &   $  6.6 ~10^{5}  $                 \\      
    \hline\hline
    
  \end{tabular}
\end{table*}

The criteria we considered for the target choice are:
\begin{itemize}
\item the maximization of the number of $\mu$ pairs produced;
\item the  minimization of the muon emittance;
\item the largest positrons survival, if needed for the positrons recirculation.
\end{itemize}
\begin{figure}[htbp!]
  \centering
  \includegraphics[width=\singlePlotSize]{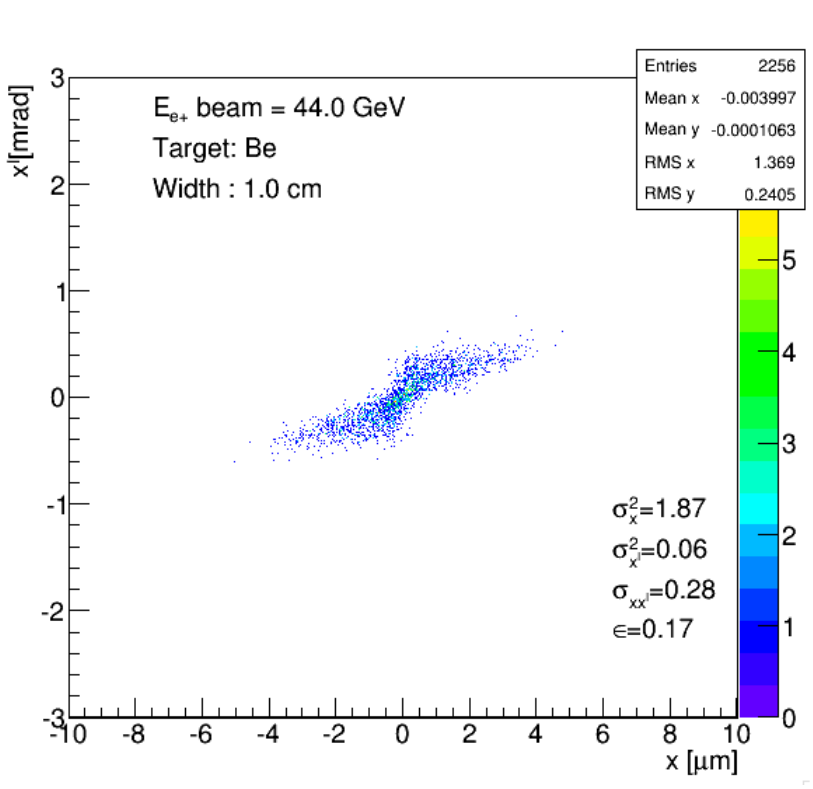}
  \caption{Horizontal phase space distribution $x-x'$ of the muons
    exiting the target, for a positron beam energy of 44 GeV and a Beryllium target of 1 cm.
    \label{fig:horPhaseSpBe}}
  \end{figure}

Positrons interactions on four different targets have
been studied with {\texttt{GEANT4}}\cite{geant}: Beryllium, Carbon, Diamond and Copper.
For these four cases, we optimized the target thickness and the positron
beam energy to maximize our key parameters. As expected, it has been found
that light materials: Beryllium, Carbon, and Diamond, have a better performance
with respect to heavier materials (i.e. Copper), having a larger muon
production efficiency $eff(\mu^+\mu^-)$. In addition in these cases the muon beam
is produced with a smaller emittance. Finally, the positron survival probability is larger for light materials. 
These characteristics can be understood looking at the values of the ratio $\lambda(\mu^+\mu-)/X_{0}$ in Table ~\ref{tab:tab1}.
Table  ~\ref{tab:simul1} shows the results of simulations performed for the positron energy of 44 GeV
for the four targets, where the thickness is chosen in order to have the same muon production rate.

\begin{figure}[htbp!]
  \centering
  \includegraphics[width=\singlePlotSize]{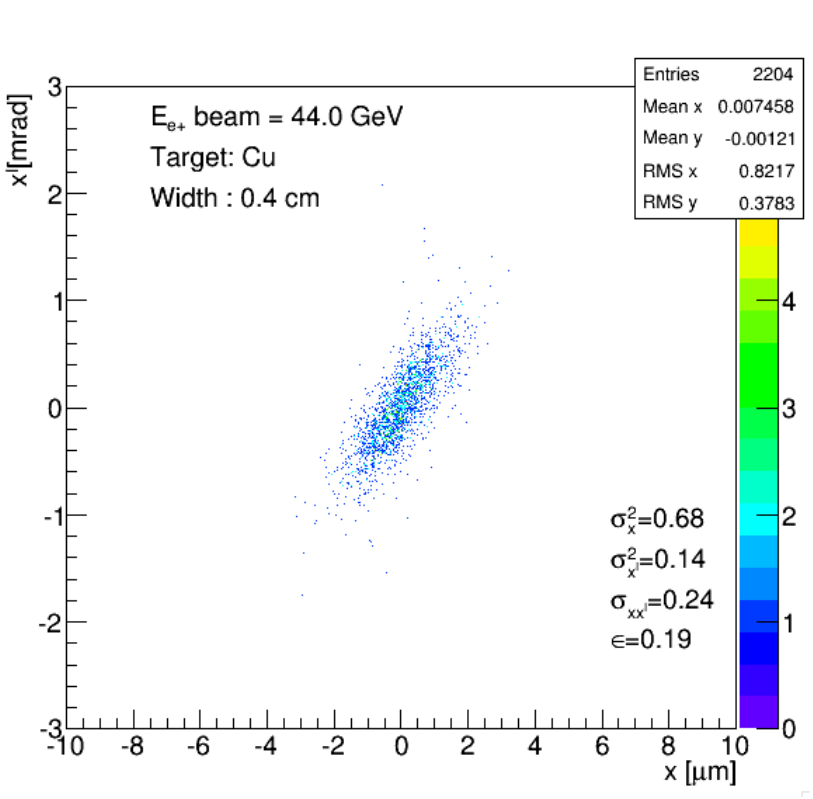}
  \caption{Horizontal phase space distribution $x-x'$ of the muons
    exiting the target, for a positron beam energy of 44 GeV and a Copper target of 1 cm.
    \label{fig:horPhaseSpCo}}
  \end{figure}

\begin{table*}[htbp!]
    \centering
  \caption{Summary of simulation results for 44 GeV positrons. \label{tab:simul1}}
  \vspace{0.05cm}
  \small
  \begin{tabular}{*{5}{l}}
    \hline\hline
        \noalign{\vspace{0.05cm}}
        &       Cu   &  C  &  Diamond  &  Be  \\
        \hline
        $L$(cm)                                &       0.4  & 0.9 &  0.5      &  1.0 \\
        $L(\lambda(\mu))  (10^{-7})$           &      2.7   &1.6  &  1.6      &  1.6 \\
        $L(X_{0})$                             &     0.29   &0.04 & 0.04      & 0.03 \\
        $\epsilon$($\mu$m-mrad)                &     0.19   &0.16 & 0.09      &      \\
        $\mu$ efficiency $(10^{-7})$           &      1.6   & 1.6 &  1.6      & 1.6  \\
        $e^+$ efficiency $(\delta E/E<10\%)$   &    0.46    &0.90 &0.90       &0.93 \\
\hline\hline
\end{tabular}
\end{table*}
The actual value of the muon production efficiency for the
Copper target is lower than that expected because of the positron
loss due to bremsstrahlung. This effect is also seen in figures~\ref{fig:horPhaseSpBe} and~\ref{fig:horPhaseSpCo}, where
it is shown the $x-x'$ distribution of the muons at the target exit. $x$ and $x'$ are the transverse
displacement and the angle with respect to the positron beam direction. The $x-x'$ distribution
for Beryllium target has the characteristic shape expected for cases in which the emittance
is dominated by target length effect and multiple scattering contributions cannot
be appreciated.  The situation is reversed for Copper target where the shape is fully dominated by multiple scattering.

\subsection{Crystal target option}
It is known that channeling phenomena are present in crystals
with particles incident angles with respect to the crystal structure
smaller than $\sqrt{2U/E}$ where $E$ is the particle energy and $U$ is the typical crystal energy
level $(O(100~\hbox{eV})$ for Diamond).  For complete channeling
there is no contribution to emittance increase due to
the target thickness and very low emittances can be obtained
with target thickness of the order of the radiation length.
For a 22 GeV muon the critical angle is about 0.1 mrad. The value of $\theta_{\mu}^{max}$ is around 0.1 mrad for $E_{+}$=43.72 GeV.
At this energy the dimuon production cross section is slightly above 0.1 $\mu$b and the muon energy
spread at 22 GeV is below 1.5\%.
We think this could be a good option in the case of an
Higgs factory at center of mass energy of 125 GeV where
a beam energy spread of about $5\cdot 10^{-5}$ is needed.

\section {Positron source \label{sec:4}}

\begin{table*}[t!]
    \centering
  \caption{Positron sources parameters for future projects from ref.~\cite{Zimmerman:2012zzb}. \label{tab:Esources}}
  \vspace{0.05cm}
  \small
  \begin{tabular}{*{6}{l}}
    \hline\hline
    \noalign{\vspace{0.05cm}}
      & SLC & CLIC & ILC & LHeC & LHeC ERL \\
    \hline
    E [GeV] & 1.19 & 2.86 & 4 & 140 & 60 \\
    $\gamma\epsilon_{x}$ [$\mu$m] & 30 & 0.66 & 10 & 100 & 50   \\
    $\gamma\epsilon_{y}$ [$\mu$m] & 2  & 0.02 & 0.04 & 100 & 50 \\
    $e^{+}$[$10^{14}$s$^{-1}$]    & 0.06 & 1.1 & 3.9 & 18 & 440 \\
    \hline\hline
  \end{tabular}
  \end{table*}

A superior positron source is required to
compensate the extremely low muon production
efficiency $eff(\mu^+\mu^-)<10^{-5}$. The present record
positrons production rate has been reached at the SLAC linac SLC. A summary
of the parameters of the positron sources for the future facilities
is reported in Table ~\ref{tab:Esources}. ILC positron source has been designed to
provide $3.9\cdot 10^{14} e^+/$s. Two order of magnitudes more intense sources are foreseen for LHeC.

\section{Muon production\label{sec:5}}

\begin{figure}[htbp!]
  \centering
  \includegraphics[width=\singlePlotSize]{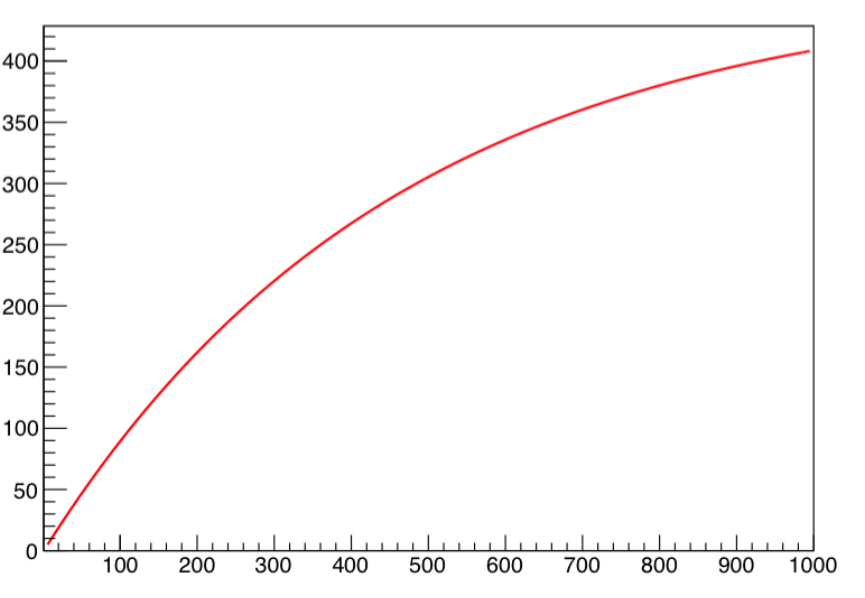}
  \caption{Muon accumulation function: number of bunches vs number of turns in the accumulator ring.
    \label{fig:muAccFunction}}
\end{figure}

In this section we present the study performed
both on the single pass and the multipass of the
positron bunch on target. In both cases the muon production rate has been maximized.
The low value of the muon conversion efficiency requires a muon
accumulator ring to reach $O(10^8)$ muons per bunch. The muons could be recombined
in two rings intercepting the positron beamline in the interaction
point of positrons on target in order to preserve the emittance.
The muon laboratory lifetime $\tau_{\mu}^{lab}$ is about 460 $\mu$s so that the recombination
of the muon bunches need to be fast. The number of bunches $n_b$  effectively accumulated
in the bunch circulating in the combiner ring at the turn $N_T$ is:
\[
n_b=\sum_{i=1}^{N_T}{e^{-\Delta t(N_T-i)/\tau_{\mu}^{lab}} }
\]
where $\Delta t$ is the positron bunch spacing equal to the muon ring revolution frequency.
Figure~\ref{fig:muAccFunction} shows the number of bunches $n_b$ as a function of
the turn number $N_T$ for $\Delta t$=1 $\mu$s; from the figure it is clear
that there is a saturation at $\sim 2\tau_{\mu}^{lab}$ and that a good
working point is around $\tau_{\mu}^{lab}$ (500 turns).

Muons accumulating in the storage ring pass the target many times and
they receive an emittance increase due to multiple scattering:

\begin{multline}
\theta_{MS}\sim\frac{1}{N_T}\sum_{i=1}^{N_T}{\frac{0.0136}{P(GeV)}\sqrt{0.5L(X_0)(N_T-i)}} \\ e^{-\Delta t(N_T-i)/\tau_{\mu}^{lab}}  
\end{multline}
\begin{multline}
X_{MS}\sim\frac{1}{N_T}\sum_{i=1}^{N_T}{\frac{0.0136}{P(GeV)}\sqrt{0.5L(X_0)(N_T-i)}}\\ \frac{0.5L}{\sqrt{3}}e^{-\Delta t(N_T-i)/\tau_{\mu}^{lab}} 
\end{multline}

We considered as a first set of parameters the number of positrons per
bunch equal to: $N_b(e^+)=4\cdot 10^{11}$, a bunch train of 2500 bunches with a
bunch spacing of 200 ns. This would give a number of positrons per bunch
train of $N_{tot}(e^+)= 10^{15}$. According to the LHeC positron rate
design, up to four bunch trains per second are feasible.
We propose that positrons at the exit of the target are collected
and conveniently reused. This scheme foresees a bunch structure that can be
obtained in an Energy Recovery Linac (ERL) with a ``single pass scheme'', or in a positron storage
ring with a ``multipass scheme''. In the following sub-sections we briefly
discuss requirements and advantages for these two schemes.

\subsection{Single pass scheme}
Single pass option requires a target with a length of
about $\sim X_0$; with such a thickness in light materials the emittance increases sizably.
For example, a simulation of positron beam of 44.5 GeV impinging on a Diamond
target of 4 cm (corresponding to 1/3 $X_0$) gives a muon efficiency $eff(\mu^+\mu^-)$  of the order
of $10^{-6}$  with an emittance of 2.5 $\mu$m-mrad.
Emittances smaller by two order of magnitudes can be obtained with crystal
targets with structures aligned to the beamline. In this case a positron
beam energy of about of 43.7 GeV has to be used and a factor of $0.3\cdot 10^{-6}$ for
the $\mu$ efficiency $eff(\mu^+\mu^-)$ is expected. Using for the positron
parameters $\sigma_{x}=0.5 ~\mu$m, $\sigma_{x'}=0.05$ mrad, an
emittance of $\epsilon_{x}=0.028~ \mu$m-mrad  can be obtained.  The total power
on target for energy loss is about 24 kW for positron beam parameters reported in the previous section.

\subsection{Multipass scheme}
A multipass scheme allows to increase the $\mu$ conversion efficiency; it can
be implemented with a large momentum acceptance  storage
ring. A 6 km positron  ring with a bending radius $\rho$ of 0.6 km has been considered. A total
positron beam current of $I_{tot}(e^+)=240$ mA, corresponding to $N_b(e^+)=3\cdot 10^{11}$  positrons
per bunch, $n_b=100$  bunches provide a rate of $1.5\cdot 10^{18}$ positrons on target per second.
Muons could be recombined in two rings with a circumference
of 60 m intercepting the positron ring in the interaction
point on target. 

\begin{table}[ht]
    \centering
  \caption{Parameters related to synchrotron emission for the positron ring. \label{tab:paramSyncEmiss}}
  \vspace{0.05cm}
  \begin{tabular}{*{2}{l}}
    \hline\hline
    \noalign{\vspace{0.05cm}}
    B [T] & 0.245 \\
    $E_{critical}$ [keV] & 315 \\
    $e^{+}$ rate [Hz] & $1.5\cdot 10^{18}$ \\
    $<N_{\gamma}>$ & 5763 \\
    $U_{0}$ [GeV] & 0.578 \\
    $P_{tot}$ [MW] & 139 \\
    \hline\hline
  \end{tabular}
\end{table}

Energy loss due to synchrotron radiation in the positron
ring has been evaluated. A summary is reported in Table ~\ref{tab:paramSyncEmiss}.  The energy loss per turn
is about 600 MeV corresponding to a total power required of about 139 MW, for a
positron rate of $1.5\cdot 10^{18}$ Hz.
The positron loss rate has to match the positron source capability. Using
LHeC positron source rate the positron loss on target has to be
below 1\%. A Beryllium target 3 mm thick provides a positron
survival probability of 2\% and 0.5\% for an energy acceptance of 5\% and 25\%, respectively.

\begin{table}[h!]
    \centering
  \caption{Energy loss in target due to bremsstrahlung. \label{tab:ElossBrem}}
  \vspace{0.05cm}
  \begin{tabular}{*{2}{l}}
    \hline\hline
    \noalign{\vspace{0.05cm}}
    length Be target [cm] & 0.3 \\
    $e^{+}$ rate [Hz] & $1.5\cdot 10^{18}$ \\
    $\Delta E/E$  (5\%)  [GeV] & 0.040 \\
    $\Delta E/E$  (25\%) [GeV] & 0.180 \\
    $P_{tot}$  (5\%)  [MW]     & 11 \\
    $P_{tot}$  (25\%) [MW]     & 48 \\
    \hline\hline
  \end{tabular}
\end{table}

The multiple scattering contribution to the emittance
amounts to 0.5 mrad  and  0.5 $\mu$m for a 3 mm Be target. Such
a thin target allows for a higher positron energy with a small 
emittance increase. At 45 GeV we obtain an emittance of 0.19 $\mu$m mrad
with a $\mu$ efficiency $eff(\mu^+\mu^-)$  of $10^{-7}$.
The muons produced with this technique have a large energy spread, being
about $\Delta E/E\approx9\%$, thus resulting interesting for high energy muon
collider and neutrino factory applications. The value of the ratio of the
number of produced muon pairs to the number of produced positrons is strongly
related to the ring energy acceptance, it is about $50\cdot 10^{-7}$ for $\Delta E/E<5\%$  and  about
 $200\cdot 10^{-7}$ for $\Delta E/E<25\%$.
Energy loss due to the radiation emitted within target has
also to be considered. For a 0.3 cm Beryllium target the power dissipated has
an increase from 3 MW to 13 MW depending on the ring energy
acceptance. The power on the target due to ionization energy is about 300 kW
requiring.

\section{Beam properties  estimate for muon collider }
\label{sec:6}
The performances of low emittance muon beams have been
studied for two cases:  multi-TeV and the Higgs factory case
at $\sqrt{s}$=125 GeV. This work has been performed to assess the
potentiality of the method; a more reliable estimate needs a 
design study to prove the feasibility of the critical issues and to 
 optimize the beams parameters.
\begin{table*}[t]
  \centering
  \caption{Comparison of muon beam properties for high energy applications
    obtained with our proposal from a positron source and with the conventional proton source.
    \label{tab:ComparisonMuBeam}}
  \vspace{0.05cm}
  \begin{tabular}{*{3}{l}}
    \hline\hline
    \noalign{\vspace{0.05cm}}
        & positron source & proton source \\
    $\mu$ rate[Hz] & $9\cdot10^{10}$ & $2\cdot 10^{13}$ \\
    $\mu$/bunch    & $4.5\cdot 10^7$ &$2\cdot 10^{12}$ \\
    normalized $\epsilon$ [$\mu$m-mrad] & 40 & 25000 \\
    \hline\hline
  \end{tabular}
  \end{table*}
\begin{table*}[t]
  \centering
  \caption{Comparison of muon beam properties for low beam energy spread
    positron sourceproton source.
    \label{tab:ComparisonMuBeamLow}}
  \vspace{0.05cm}
  \begin{tabular}{*{3}{l}}
    \hline\hline
    \noalign{\vspace{0.05cm}}
        & positron source & proton source \\
    $\mu$ rate[Hz] & $2\cdot10^{9}$ & $2\cdot 10^{13}$ \\
    $\mu$/bunch    & $10^9$ &$2\cdot 10^{12}$ \\
    normalized $\epsilon$ [$\mu$m-mrad] & 0.059 & 25000 \\
    \hline\hline
  \end{tabular}
  \end{table*}

\subsection{Multi-TeV case}
We consider that $\mu^+$ and $\mu^-$ beams are produced, as
described in section 4.2, from a 45 GeV positron beam impinging
on a 3 mm Beryllium target. We considered $3\cdot10^{11}$ positrons per bunch
with 100 bunches that circulate in a 6 km positron ring with an energy
acceptance as large as $\pm 5\%$. The muon bunches that are produced by the
positron beam are accumulated in two separate
combiner rings, one for $\mu^+$ and one for $\mu^-$, with a circumference of 60 m and circulating for 2500 turns.

The muon collider ring would have bunches of $\mu^+$  and $\mu^-$ with
energy of 22 GeV with $4.5\cdot 10^7$ muon particles, emittance
0.19 $\mu$m-mrad, and beam energy spread of 9\%, produced with a spacing of
500 $\mu$s (2 KHz rate). Bunches can be accelerated to
the nominal energy as studied by the Muon Accelaration Program (MAP) working group~\cite{map:map}. 

The relevant parameters needed to determine the luminosity in our
proposal of muon collider are reported in table 6. These performances
can be compared with those reported in Ref.~\cite{map:map},  also shown in table ~\ref{tab:ComparisonMuBeam}. From this
table it is clear that the quality of the muons produced from a positron
source as we propose in this paper is much better than the
one obtainable with a proton source; however, the muons rate is a key parameter.
We think that further studies are needed to set a maximum limit in our scheme.

Promising values of luminosities can be obtained with these parameters, being
in the range of $L\approx 10^{32}$ cm$^{-2}$s$^{-1}$.

\subsection{Muon collider at the Higgs boson energy: $E_{cm}$ = 125 GeV}
The optimal scheme for a muon Higgs factory collider is with a
single pass scheme with an interaction of positron beam on target
just above the dimuon threshold. We propose a positron beam structure
similar to the ILC one, with 500 $\mu$s long bunch trains.
Each bunch train contains 2500 bunches spaced by 200 ns providing a total
rate of positrons of $6\cdot10^{15}$ Hz.
The natural muon beam energy spread is about 0.04\% at 62.5 GeV for
a positron beam of 43.8 GeV. It might be reduced to the required values
with an increase of the bunch length of 50 times.
At these positrons energies a crystal target can be used to obtain a very
low emittance. In a 4 cm Diamond crystal a muon conversion efficiency
of $3\cdot10^{-7}$ with an emittance  of 0.028 $\mu$m-mrad can
be obtained with muon rates of $2\cdot10^9$ Hz. Comparison with proton
source results from MAP is given in table~\ref{tab:ComparisonMuBeamLow}.

\section{Conclusion}
We have presented a novel scheme for the production of muons starting
from a positron beam on target, discussing the critical aspects and
key parameters of this idea and giving a consistent set of possible
parameters that show its feasibility. This scheme has several
advantages, the most important one is that it solves the problem of
muon cooling. In fact, muon beams are generated already with very low emittances i.e.
comparable to that obtained with electron beams. In addition, it might
be able to provide luminosity with low muon fluxes avoiding the problems
of irradiation typical with the conventional proposal.
A critical point is the production of the necessary muon rate that 
requires detailed studies to assess the maximum possible value. First
results presented in this paper shows that first class positron sources
proposed for ILC and LHeC need are marginally sufficient to this purpose. An improvement
 in the positron rate is required for a muon collider purpose. Target survival needs also deep studies. 
A first set of parameters for a muon collider at high energy and 125 GeV has been
shown to assess the potentiality of this proposal.
We think that the promising results discussed in this paper
encourage a serious design study of the proposal.

\end{document}